%% file: main.tex
\pdfoutput=1
\documentclass[acmsmall,screen,nonacm]{acmart}
\usepackage{graphicx} 
\usepackage{xcolor}
\usepackage{pifont}
\usepackage{multirow}
\usepackage{makecell}
\usepackage{tcolorbox}
\newtcolorbox{findingbox}[1]{colback=yellow!10!white,colframe=yellow!50!black,title=#1}
\usepackage{enumitem}
\usepackage{booktabs}
\usepackage{subcaption}
\usepackage[utf8]{inputenc}
\usepackage{textcomp}

\usepackage{algorithm}
\usepackage{algpseudocode}

\usepackage{listings}
\DeclareUrlCommand\rulename{%
  \urlstyle{tt}%
}
\usepackage{comment}

\newif\ifshowplan
\showplantrue
\ifshowplan
    {\par\color{blue!70!black}\small\begin{itemize}\setlength\itemsep{1pt}}%
    {\end{itemize}\par}
\else
  \excludecomment{plan}
\fi

\newcommand{\finding}[2]{%
  \par\medskip\noindent
  \fbox{\parbox{\dimexpr\linewidth-2\fboxsep-2\fboxrule\relax}{%
    \textbf{Finding #1.} #2}}%
  \par\medskip}
  
  \newcommand{\twoline}[2]{\begin{tabular}[b]{@{}c@{}}\textbf{#1}\\\textbf{#2}\end{tabular}}
  
\usepackage{adjustbox}

\newcommand{\rqone}{\textbf{RQ1}}
\newcommand{\rqtwo}{\textbf{RQ2}}
\newcommand{\rqthree}{\textbf{RQ3}}

\begin{document}

\title{Static Analysis Guidance for Secure LLM Code Generation}

\author{Daniele Lunghi}
\authornote{The first three authors contributed equally to this work.}
\affiliation{%
  \institution{University of Luxembourg}
  \city{Luxembourg}
  \country{Luxembourg}
}
\email{daniele.lunghi@uni.lu}

\author{Alberick Euraste Djire}
\authornotemark[1]
\affiliation{%
  \institution{University of Luxembourg}
  \city{Luxembourg}
  \country{Luxembourg}
}
\email{euraste.djire@uni.lu}

\author{Zacharie Chenail-Larcher}
\authornotemark[1]
\affiliation{%
  \institution{École de technologie supérieure (ÉTS)}
  \city{Montréal}
  \state{QC}
  \country{Canada}
}
\email{zacharie.chenail.larcher@gmail.com}

\author{Melissa Tessa}
\affiliation{%
  \institution{University of Luxembourg}
  \city{Luxembourg}
  \country{Luxembourg}
}
\email{melissa.tessa@uni.lu}

\author{Quentin Stiévenart}
\affiliation{%
  \institution{Université catholique de Louvain}
  \city{Louvain-la-Neuve}
  \country{Belgium}
}
\email{stievenart.quentin@uqam.ca}

\author{Jacques Klein}
\affiliation{%
  \institution{University of Luxembourg}
  \city{Luxembourg}
  \country{Luxembourg}
}
\email{jacques.klein@uni.lu}

\author{Florent Avellaneda}
\affiliation{%
  \institution{Université du Québec à Montréal}
  \city{Montréal}
  \state{QC}
  \country{Canada}
}
\email{avellaneda.florent@uqam.ca}

\author{Naouel Moha}
\affiliation{%
  \institution{École de technologie supérieure (ÉTS)}
  \city{Montréal}
  \state{QC}
  \country{Canada}
}
\email{Naouel.Moha@etsmtl.ca}

\author{Tegawendé F. Bissyandé}
\affiliation{%
  \institution{University of Luxembourg}
  \city{Luxembourg}
  \country{Luxembourg}
}
\email{tegawende.bissyande@uni.lu}

\input{sections/abstract}

\maketitle
\renewcommand{\shortauthors}{Lunghi, Djire, Chenail-Larcher et al.}

\input{sections/introduction}

\input{sections/background}

\input{sections/study_design}

\input{sections/results}

\input{sections/related_works}

\input{sections/discussion}

\input{sections/threath_of_validity}

\input{sections/conclusion}

\bibliographystyle{ACM-Reference-Format}
\bibliography{references}

%


\end{document}

\end{document}

%% file: sections/abstract.tex
\begin{abstract}
Static analyzers are cheap and widely deployed, which makes their reports a candidate feedback signal for secure LLM code generation. Existing approaches place the analyzer inside an iterative repair loop with intermediate steps, which is costly and makes it hard to isolate the report's effect on the model. We introduce two strategies that give the finding to the model during generation. When the analyzer raises an alert, generation is rolled back, and the finding is inserted either as a code comment (Comment) or in the user message (Steering). We compare them with two iterative approaches and a control without the analyzer, across three models and three Python benchmarks, and we measure security using each benchmark's oracle.
Guidance during generation is cheap and can be effective, at 0.9 to 1.9 times the baseline’s output tokens against up to 24.1 times for iterative approaches. With DeepSeek-V4.1-flash on RiskyPy, Steering raises the share of functional and secure programs from 52.5\% to 64.4\%, close to the best iterative approach (66.2\%) while using less than a fifth of the tokens. A per-rule analysis shows that rules with a local fix benefit most, and replacing the XML parser is net-positive on two of the three models. The study also identifies when guidance should not be applied. Analyzers flag at most 5.6\% of the functional but insecure programs on SecEvalPlus, leaving little guidance to act on. When a static analysis alert (warning or error) contradicts the task, models tend to follow the analyzer and break the program. The two Qwen2.5 models often fail when the fix requires restructuring the code.  These results indicate that static-analysis guidance is worth applying with a capable model, on rules with a local fix, and after filtering alerts that conflict with the task specification.
\end{abstract}

\keywords{Large language models, code generation, software security, static analysis, empirical study}

%% file: sections/introduction.tex
\section{Introduction}
\label{sec:intro}

Large Language Models (LLMs) now write a growing share of test and production code~\cite{gao2025current, jiang2026survey, dong2025survey}. However, the code they generate is often insecure~\cite{wang2024your}. Even frontier models produce many backend applications that are incorrect or exploitable~\cite{vero2025baxbenchllmsgeneratecorrect}, and comparisons with human-written code show that LLM-generated code exhibits its own types of vulnerabilities~\cite{cotroneo2025human}. As generated code is integrated into continuous integration pipelines with less human review, automated security checks become necessary, and their results need to be fed back to the model.

Static analysis is a well-established way to perform such checks. Static analyzers examine source code without executing it, which makes them faster and cheaper than dynamic analysis or formal verification~\cite{thomson2021static}. They are mature, widely deployed~\cite{zheng2006value}, and integrated into IDEs, code review, and CI pipelines. Each issue reported by an analyzer, which we refer to as an \emph{alert}, is associated with a line of code and a rule, which makes it easy to provide as feedback to a model. However, static analyzers have well-known limitations. A rule matches a code pattern regardless of whether the task requires it, which leads to many false positives that developers tend to ignore~\cite{christakis2016developers, johnson2013why}. In addition, static analyzers mainly detect local, syntactic weaknesses and miss most logic, authentication, configuration, and runtime flaws~\cite{zheng2006value}. Imperfect feedback has nevertheless been used successfully to guide LLMs. For example, unit tests are used to filter generated code~\cite{peng2025perfcodegen} and compiler errors are used for self-repair~\cite{dou2024stepcoder}. It remains unclear whether static-analysis alerts can play the same role.

Prior work has used static analyzers to secure generated code, but only within iterative repair loops. PromSec~\cite{nazzal2024promsec} uses a graph neural network and an LLM to rewrite the prompt, and FDSP~\cite{alrashedy2024llmspatchsecurityissues} asks an LLM to turn the analyzer report into candidate fixes. In both approaches, the alert goes through intermediate steps before it reaches the model, so its effect on the generated code cannot be isolated. In this paper, we propose two strategies that pass the alert to the model during generation. A harness runs the analyzer after each generated line. When the analyzer raises an alert, the harness rolls back the code to before the flagged line and provides the alert to the model, either as a comment in the code (\emph{Comment}) or in the user message (\emph{Steering}). Only the code after the rollback point is regenerated.

We compare these two strategies with PromSec, FDSP, an unguided baseline, and Self-Refine, a control that uses no analyzer. We conduct the comparison on three LLMs from two families and on three Python benchmarks. We select the benchmarks according to how well alerts coincide with real insecurity and with the task specification, which we refer to as the \emph{static-analysis alignment spectrum}. RiskyPy~\cite{riskypy} tasks were built so that analyzers raise alerts on their solutions, SecEvalPlus~\cite{wang2024your} solutions rarely trigger alerts, and on BaxBench~\cite{vero2025baxbenchllmsgeneratecorrect} the most frequent alert contradicts the specification. To avoid a circular evaluation, we measure security with the oracle of each benchmark and not with the analyzers used for guidance. These oracles are security tests for SecEvalPlus, live exploits for BaxBench, and an LLM judge validated against human annotations for RiskyPy. We also measure functionality and cost. Finally, we classify the outcome of each alert raised during Steering as \emph{Solved}, \emph{Broke}, \emph{Did nothing}, \emph{Already broken}, or \emph{Silenced} (the alert is cleared, but the program is still insecure).

Our results show that guidance during generation is inexpensive and that it is effective under specific conditions. Comment and Steering require 0.9 to 1.9 times the output tokens of the baseline, whereas the iterative approaches require up to 24.1 times. With DeepSeek-V4.1-flash on RiskyPy, Steering increases the share of functional and secure programs from 52.5\% to 64.4\%. FDSP and Self-Refine reach 66.2\% and 63.5\%, but they require 9.8 and 18.3 times the baseline tokens, whereas Steering requires 1.7 times. The per-rule analysis shows that rules with a local fix benefit most. For example, the rule that requires replacing the XML parser has a net-positive effect on two of the three models.

However, the gains are not uniform across models and benchmarks, and no strategy improves on the baseline on all three benchmarks. We identify three conditions under which guidance is effective. First, the analyzer must be able to detect the weakness. On SecEvalPlus, Semgrep and Bandit flag at most 5.6\% of the functional but insecure programs. Second, the alert must be compatible with the task. When an alert contradicts the task, models tend to follow the analyzer. We refer to this behavior as \emph{overcompliance}. On BaxBench, the models removed a network binding required by the specification in 79 of 83 cases, although the prompt instructed them to give priority to the task. Third, the model must be capable of applying the fix. The two Qwen2.5 models often break the program when the fix requires restructuring the code. We also observe that analyzer alerts are not a reliable measure of security gains. On RiskyPy, the analyzer-guided strategies significantly increase the share of functional programs without alerts in 12 of 12 comparisons, whereas the benchmark oracle confirms a significant security gain in only 2 of them.

In summary, this paper makes the following contributions.
\begin{itemize}
\item We introduce two black-box strategies, Comment and Steering, that pass analyzer alerts to the model during generation without intermediate steps. They require 0.9 to 1.9 times the output tokens of the baseline.
\item We conduct, to our knowledge, the first systematic study of static-analysis guidance for LLM code generation on benchmarks that span the alignment spectrum. The study covers security, functionality, and cost, and it measures security with oracles that are independent of the guiding analyzers.
\item We analyze the outcome of each alert per rule. We show that rules with a local fix benefit from guidance, and we identify \emph{overcompliance} as a failure mode when an alert contradicts the task.
\item We show that evaluating analyzer-guided generation with the guiding analyzer overstates its security benefit.
\item We release a replication package with code, prompts, generations, human labels, and evaluation scripts.
\end{itemize}

%% file: sections/background.tex
\section{Background}
\label{sec:background}
This section presents the two topics our study builds on, namely static analysis for security and the use of feedback to guide LLM code generation.
\subsection{Static Analysis for Security}

Static Application Security Testing (SAST) analyzes source code without executing it in order to identify potential vulnerabilities, which are usually mapped to Common Weakness Enumeration (CWE) classes such as OS Command Injection (CWE-78) or SQL Injection (CWE-89). Because it requires neither a runtime environment nor test inputs, SAST can be applied early and is often integrated into CI pipelines to support code review~\cite{sadowski2015tricorder}. For LLM-generated code, SAST tools are used both to evaluate the security of the code~\cite{zhao2026vibecodingsafebenchmarking} and to provide feedback in iterative refinement loops~\cite{alrashedy2025leveraging}.

Static analyzers differ in precision and cost. Bandit~\cite{bandit2024} builds the AST of a Python program and runs security plugins on it. Semgrep~\cite{semgrep} applies declarative rules that follow the syntax of the source code and are mapped to CWE classes. More advanced analyzers such as CodeQL~\cite{codeql} store the code in a relational database and perform data-flow and taint analysis across functions.

Static analyzers have two main limitations. First, their precision is limited. Most tools have high false-positive rates and give little context on the reported issues~\cite{johnson2013why, christakis2016developers}, and on LLM-generated code the verdicts of Semgrep and CodeQL frequently diverge from expert annotations~\cite{firouzi2026persistent}. A main cause is that analyzers do not consider the context of the application. \citet{alecci2026damflow} show that even mature taint analyzers such as FlowDroid produce false alarms and false negatives because fixed lists of sources and sinks do not account for the semantics of the application domain. Second, their scope is limited. Pattern-based rules cannot detect business logic flaws, missing authorization checks, or vulnerabilities that depend on runtime behavior.

\subsection{Feedback-Guided Code Generation for Security}

A common way to secure LLM-generated code is to evaluate the code after generation and to feed the result back to the model, which revises the code until no problem is reported or a budget is reached. The same loop is used for correctness, with test execution and compiler errors as feedback~\cite{chen2023teachinglargelanguagemodels}. Following \citet{liu2025survey}, we distinguish \emph{internal} feedback, produced by the model itself, from \emph{external} feedback, produced by an independent tool such as a static analyzer. Other paradigms for securing generated code~\cite{basic2026vulnerabilitiesremediationsystematicliterature} modify the model through training, as in prefix control~\cite{he2023large} and instruction tuning~\cite{he2024instruction}, or act on the logits during decoding, for security~\cite{fu2024constrained} or for correctness~\cite{agrawal2023monitor}. These approaches require white-box access. We consider only black-box approaches, which also apply to models served through an API. An intervention during generation can be black-box if it interrupts generation and modifies the input of the model before generation resumes, which is the principle of the two strategies we propose.

\textbf{Internal feedback.} Self-Refine~\cite{madaan2023self} asks the model to critique its own output and to revise it. It needs no external tool but relies on the judgment of the model, and each round rewrites the whole program. \citet{yan2025guiding} report that self-generated hints help only when the model follows instructions well, and that weaker models benefit little even from high-quality hints.

\textbf{External feedback.} Analyzer-based approaches generate code, analyze it, and use the alerts to produce a new version, until the analyzer reports nothing or a budget is reached. They differ in how directly the report reaches the model. In PromSec~\cite{nazzal2024promsec}, the model never receives the report. A graph-guided GAN perturbs a graph representation of the code, an LLM rewrites the code according to this graph, and Bandit~\cite{bandit2024} is used only as the objective and the stopping condition. In FDSP~\cite{alrashedy2024llmspatchsecurityissues}, the model receives the code and the Bandit report, proposes several candidate fixes, and refines the code with each of them until Bandit reports nothing. Other approaches add knowledge beyond the analyzer, such as retrieved security examples or secure-coding guidelines~\cite{patir2025fortifying, saul2025scgagent}. We exclude them because this additional knowledge would confound the effect of the analyzer feedback.

%% file: sections/study_design.tex
\section{Study Design}
\label{sec:design}

\subsection{Research Questions}
\label{sec:rqs}

Our study addresses the following research questions (RQs).

\begin{description}
    \item[\rqone{} (Scope):] To what extent can static analysis assess the security of LLM-generated code?
    \item[\rqtwo{} (Strategies):] How do guidance strategies compare in security, functionality, and cost?
    \item[\rqthree{} (Factors and failures):] Which factors determine whether static-analysis guidance helps, and why does it fail?
\end{description}

\rqone{} establishes an upper bound for guidance, since guidance can only address the weaknesses that the analyzer reports. We measure how many insecure programs the analyzers flag and how often flagged programs are insecure (Section~\ref{sec:rq1}). \rqtwo{} compares the strategies in terms of security, functionality, and cost against an unguided baseline and Self-Refine (Section~\ref{sec:rq2}). \rqthree{} explains the observed differences by analyzing the per-rule outcomes of Steering with respect to the model, the rule, and the benchmark (Section~\ref{sec:rq3}).

\subsection{Guidance Strategies}
\label{sec:strategies}

We select three existing approaches that differ in how much they rely on the static analyzer (Section~\ref{sec:background}). Self-Refine~\cite{madaan2023self} uses no analyzer and serves as a control, which separates the effect of iterating on the code from the effect of the analyzer's feedback. PromSec~\cite{nazzal2024promsec} uses Bandit only as its objective and stopping condition, and FDSP~\cite{alrashedy2024llmspatchsecurityissues} provides the Bandit report to the model.

In addition, we propose two strategies that guide the model during generation. Both follow the same principle. When the analyzer raises an alert on the code being generated, the harness rolls back the flagged code and modifies the model's context before generation resumes.

\subsubsection{Comment}

Comment guides the model through comments inserted in the generated code. The system prompt asks the model to write, before each logical block of code, a comment of the form ``\texttt{\# We will now \ldots}'' that describes what will be implemented next. We use these \emph{intent comments} as rollback points.

During generation, the harness runs the analyzer each time the model completes a line, as long as the partial code is syntactically valid. We use Semgrep with the \texttt{p/security-audit} rule set. When the analyzer raises an alert, generation stops and the code is rolled back to the last intent comment before the flagged line. If several alerts are raised, we use the earliest one in the code. The harness then inserts, after the intent comment, a comment of the form ``\texttt{\# I will avoid this issue: \textless{}message\textgreater{}}'' that contains the analyzer's message, and generation resumes from this point. Since only the code after the intent comment is regenerated, Comment requires in principle fewer additional tokens than regenerating the whole program.

\subsubsection{Steering}

Steering guides the model through its prompt. Unlike Comment, it does not modify the generated code. The harness runs the analyzer as in Comment. When an alert is raised, generation stops and the harness removes the flagged line and all subsequent lines. The alert is then added to a dedicated section at the end of the user message, which is the part of the prompt closest to the code being generated. For each alert, this section contains the rule identifier, its CWE, the analyzer's message, and a recommendation for the rule. The prompt instructs the model to follow the recommendation only if it does not conflict with the task requirements. We generated the recommendations before the experiments, with \texttt{gpt-5.2-2025-12-11}, for each Python rule of the \texttt{p/security-audit} rule set.

If the same alert is raised again in the next round, the code is rolled back to the beginning of the generation. This full rollback allows the model to revise an early decision that a local rollback cannot correct. For example, once a model has imported a package, it tends to use it.

The dedicated section contains at most two alerts, and a new alert replaces the oldest one when the section is full. A sample is rolled back at most five times. The process also stops when a round would repeat an earlier prompt and code, which prevents infinite loops.

\subsubsection{Implementation Adjustments}
\label{sec:adjustments}

We run all strategies with the same harness, the same system prompt, and greedy decoding. We adapt the existing approaches only when this is required to run them or to ensure a fair comparison, and we document every change in the replication package. All strategies are limited to five rounds. Semgrep uses a frozen snapshot of \texttt{p/security-audit} (225 rules) and reports alerts of ERROR and WARNING severity. Bandit reports alerts of medium or higher severity and confidence.

\paragraph{Steering.} When Steering exhausts its rollbacks or repeats a round while the alert is still present, the truncated code may be incomplete. In this case, the model completes the code once with the analyzer disabled, the guidance still in the prompt, and a comment that names the unresolved alert. The resulting program is thus runnable, and the per-rule analysis labels the alert as \emph{Did nothing}.

\paragraph{FDSP} We follow the authors' implementation. The model verbalizes the Bandit report, proposes candidate fixes, and repairs the code with each candidate until Bandit reports no alert. We implement the early stop described in the paper, discard candidates that do not parse, since an analyzer reports nothing on them, and return the candidate with the fewest alerts instead of the last one. To separate the effect of the analyzer from the effect of the strategy, we also evaluate \textit{FDSP (Semgrep)}, where Semgrep replaces Bandit while keeping the rest of the pipeline unchanged.

\paragraph{PromSec.} The authors' release does not include a trained generator. We therefore train the gGAN with the authors' procedure and hyperparameters, on an AST-based control-flow graph with per-node features, and we pass each node's source line and features to the LLM as text. As for the other strategies, the initial code is generated from the plain task. The objective remains the number of distinct CWEs reported by unfiltered Bandit. We return the best-scoring round instead of the last one, which slightly favors PromSec.

\paragraph{Self-Refine.} We use the standard generate-critique-refine loop, which stops when the critique reports no issue.

\subsection{Datasets}
\label{sec:datasets}

We require benchmarks that \ding{172} elicit potentially vulnerable code, \ding{173} provide a functional oracle, and \ding{174} assess security independently of the analyzers used for guidance. In addition, we select benchmarks according to how well alerts coincide with real insecurity and with the task specification, which we refer to as the \emph{static-analysis alignment spectrum}. We select one benchmark built so that analyzers raise alerts on its solutions, one where alerts are rare, and one where the most frequent alert contradicts the specification.

\paragraph{SecEvalPlus.} CodeSecEval~\citep{wang2024your} targets the CWE Top~25 and provides functional tests and security tests for each problem. We use its SecEvalPlus subset (140 code-synthesis tasks) because our study targets generation from a specification and not code completion. Many of the targeted weaknesses have no syntactic pattern that an analyzer can match. As a result, analyzers rarely raise alerts on its solutions, and guidance is rarely triggered.

\paragraph{BaxBench.} BaxBench~\citep{vero2025baxbenchllmsgeneratecorrect} is a multi-language benchmark of backend applications, evaluated with functional tests and end-to-end exploits against the running server. We keep the 84 Python tasks that require a single generated file, because all the studied strategies process one code snippet at a time. Each application runs in a container, and the specification requires the server to listen on \texttt{0.0.0.0} so that the tests can reach it. The Semgrep rule \texttt{avoid\_app\_run\_with\_bad\_host} flags this binding. A model that follows the alert typically binds the server to a loopback address, which makes the application unreachable. In this case, the functional tests fail and the exploits cannot reach the server, which lowers the attack rate without making the code more secure. BaxBench therefore represents the case where the most frequent alert contradicts the specification.

\paragraph{RiskyPy.} RiskyPy~\citep{riskypy} contains 73 Python tasks that cover 18 CWEs of the 2025 CWE Top~25. A task was kept only if a small canary model solved it functionally and Bandit or Semgrep flagged the solution of a stronger canary model. RiskyPy is therefore aligned with static analysis by construction and is the most favorable setting for guidance. We assess functionality with the benchmark's LLM judge. Since measuring security with the analyzers would be circular, we assess security with an LLM judge that we validate against human annotations (Section~\ref{sec:human-eval}).

\subsection{Models}
\label{sec:models}

We evaluate three open-weight models from two families. From the Qwen2.5 family, we select Qwen2.5-Coder-14B-Instruct~\cite{hui2024qwen25codertechnicalreport} and Qwen2.5-72B-Instruct~\cite{qwen2025qwen25technicalreport}, which allows us to study the effect of model size. We also select DeepSeek-V4.1-flash~\cite{xu2026deepseek}, a substantially larger model from a different provider. Qwen2.5-Coder-14B runs locally on a server with one NVIDIA RTX 5000 (32\,GB), an Intel Xeon Silver 4416+ CPU, and 25\,GB of RAM. DeepSeek-V4.1-flash is queried through the official DeepSeek API and Qwen2.5-72B through OpenRouter.

All experiments use greedy decoding (temperature~$0$) and a maximum output length of 4096 tokens. Local inference produced identical outputs across repeated runs, so we run each Qwen2.5-Coder-14B experiment once. The API-served models are not deterministic even at temperature~$0$. We therefore run each of their experiments three times and report the mean.

\subsection{Metrics and Outcome Taxonomy}
\label{sec:metrics}

\paragraph{Security and functionality.} We evaluate the final program of every strategy, task, and model with each benchmark's oracles (Section~\ref{sec:datasets}). On SecEvalPlus, a program is functional if it passes the functional tests (Test-FP) and secure if it passes the security tests (Test-SP). On BaxBench, following the original protocol~\citep{vero2025baxbenchllmsgeneratecorrect}, each application runs in a Docker container. It is functional if it passes all functional tests and insecure if at least one exploit succeeds. On RiskyPy, the original LLM judge~\citep{riskypy} (\texttt{gpt-5.1-2025-11-13}, medium reasoning effort) considers a program functional if it would run without errors and accomplish the task. We use the same judge and settings for security, where a program is secure if it contains no vulnerability that can be exploited in the context of the task. Our main metric is the \emph{combined rate}, i.e., the share of tasks whose program is both functional and secure. All proportions are computed over all tasks and, for the API models, averaged over the three runs.

\paragraph{Analyzer-based metrics.} We also run Semgrep (same snapshot, ERROR and WARNING severity) and Bandit (medium or higher severity) on every final program. A program is \emph{flagged} if an analyzer raises at least one alert. We report each analyzer's flag rate and the share of programs that are functional and flagged by neither analyzer. For \rqone{}, we compare flags with the oracle on functional programs. \emph{Recall} is the share of insecure programs that are flagged, and \emph{precision} is the share of flagged programs that are insecure.

\paragraph{Statistical tests.} We compare each strategy with the baseline on the combined rate and on the share of programs that are functional and flagged by neither analyzer. Comparing the two metrics shows whether fewer alerts correspond to more secure programs. We use exact McNemar tests on task outcomes paired with the baseline program for the same task and run. For the API models, we combine the three runs with a Cochran--Mantel--Haenszel test for matched pairs, stratified by run. We apply the Holm correction within each model, benchmark, and metric.

\paragraph{Cost.} For each strategy, we report the number of output tokens over all model calls, including the tokens discarded by a rollback. 
We also report wall-clock time, which is only indicative because it depends on server load and API latency.

\paragraph{Per-rule outcomes.} We analyze the Steering's outcome per rule. For each sample generated with Steering, we consider every Semgrep rule that interrupted generation at least once. The unit of analysis is a pair of a sample and such a rule, and $n$ denotes the number of pairs. A rule \emph{fires} on a sample when Semgrep raises at least one alert of this rule on the final code. We compare each pair with the baseline sample for the same task and run, and assign the first of the following outcomes that holds. The benchmark's oracle determines security.

\begin{description}
\item[Did nothing:] The code is still detected as vulnerable by the rule, or the baseline code was already secure.
\item[Broke:] The steered sample is not functional while the baseline sample is.
\item[Already broken:] The code was already broken in the baseline. We report these pairs but do not interpret them.
\item[Silenced:] The rule no longer fires and the steered sample is functional, but it is still insecure.
\item[Solved:] The rule no longer fires, and the steered sample is functional and secure.
\end{description}

Functionality and security are properties of the whole sample, so every pair inherits its sample's verdicts. If a fix breaks the program, all the rules of the sample are labeled Broke. If the functionality remains but an unrelated weakness remains, every rule that no longer fires is labeled Silenced. The Broke and Silenced rates are therefore upper bounds for the true values, and the Solved rate is a lower bound. Solved requires the baseline sample to be insecure, as it refers only to cases where the approach solved all security issues from a functional baseline generation without breaking it. Silenced does not necessarily mean that the vulnerability was hidden.

\subsection{Validating the RiskyPy Judge}
\label{sec:human-eval}

RiskyPy was built so that static analyzers flag the solutions of its tasks, and several strategies are guided by the same analyzers. Measuring security with these analyzers would therefore be circular, which is an instance of Goodhart's law~\cite{manheim2018categorizing}. We assess security with an LLM judge and validate it through a manual annotation study.

Two software engineering experts, with 8 and 7 years of experience, independently labeled the same 50 randomly selected RiskyPy generations as secure or not secure, using the definition given to the judge. The two annotators agree on 33 of the 50 generations (66.0\%, $\kappa=0.361$), which corresponds to fair agreement on the Landis and Koch scale~\cite {landis1977}. The judge agrees with each annotator on 74.0\% and 72.0\% of the generations ($\kappa=0.463$ and $\kappa=0.467$), which corresponds to moderate agreement. In the 33 generations where the annotators agree, the judge gives the same label in 84.9\% of cases ($\kappa=0.699$), corresponding to substantial agreement.
The judge thus agrees with each annotator more often than the annotators agree with each other. The annotators differ mainly in strictness. One labeled 30 generations as secure and the other 17, and in 15 of their 17 disagreements the first annotator considered the generation secure and the second did not. On these 17 generations, the judge agrees with one annotator 9 times and with the other 8 times. Its labels therefore fall within the range of disagreement between the two annotators.

%% file: sections/results.tex
\input{sections/results/scope_of_analysis}

\input{sections/results/guidance_comparaison}

\input{sections/results/drive_success}

\input{sections/results/failure_mode}

%% file: sections/results/scope_of_analysis.tex
\section{Results}
\subsection{RQ1: Scope of Static-Analysis Guidance}
\label{sec:rq1}
Guidance can only act on what the analyzer reports. We therefore compare the analyzers' flags with each benchmark's security oracle on the baseline's functional programs, using the recall and precision defined in Section~\ref{sec:metrics} (Table~\ref{tab:analyzer-precision-recall}). Recall bounds what guidance can achieve, since an unflagged weakness gives the model nothing to fix. Precision measures how often guidance targets a real problem rather than a false alarm that invites the model to change working code for no security gain.


\paragraph{Recall: most insecurity is invisible.} On SecEvalPlus, insecurity is common but almost invisible to the analyzers. Between 62.0\% and 87.3\% of functional programs fail at least one security test, yet the analyzers flag only 4.2-5.6\% of them. Even a strategy that fixed every flagged weakness perfectly would leave more than 94\% of insecure functional programs untouched. On BaxBench, Semgrep flags 32.4\% of the exploitable applications generated by DeepSeek-V4.1-flash. On RiskyPy, whose tasks were selected because analyzers flag their solutions, recall reaches 80.0--93.8\%, yet between 6\% and 20\% of insecure programs remain outside the analyzer's view. Recall is moreover an optimistic ceiling. A flagged program may contain further weaknesses the analyzer does not report, so removing the flagged pattern does not necessarily make it secure.

\paragraph{Precision: flags are weakly informative.} A flag is informative only if flagged programs are insecure more often than functional programs in general. On SecEvalPlus and BaxBench, this is barely, if at all, the case. For DeepSeek-V4.1-flash on BaxBench, 38.6\% of flagged applications are exploitable, against 41.0\% of all functional applications: the flag carries no information about exploitability. Qwen2.5-72B shows a modest lift (68.0\% against 61.2\%), and Qwen2.5-Coder-14B produces too few functional applications ($N = 12$) to interpret. On SecEvalPlus, precision for both Qwen models (80.0--83.3\%) does not exceed the prevalence of insecurity (81.1--87.3\%). Only DeepSeek-V4.1-flash shows a modest lift (75.0\% against 62.0\%) on few flagged programs.

RiskyPy's high recall does not make its flags informative either. The analyzers flag 75.5--89.6\% of all functional programs, secure or not, so precision (47.6--61.2\%) barely exceeds the base rate of insecurity (45.5--58.0\%), a lift of only about $1.05\times$. Between 39\% and 52\% of flagged functional programs are judged safe.

\finding{1}{Static analysis sees only a small part of the insecurity in LLM-generated code, and its flags are barely more likely than chance to mark insecure code. Where analyzers rarely fire, most insecurity goes unseen. Where they fire often, they flag secure and insecure code alike. Analyzer-driven guidance must therefore contend with many undetected issues and a high false-alarm rate.}

\begin{table*}[t]
\centering
\small
\caption{Analyzer flags as a predictor of insecure code, on the Baseline's functional code only. Functional / insecure: BaxBench: passes the functional tests / an exploit succeeded; SecEvalPlus: passes Test-FP / fails Test-SP; RiskyPy: the LLM judge says it works / the LLM judge says insecure. N: functional samples with a verdict. Insecure, Flagged: \% of N. Precision and recall of the flag for insecure code, in \%.}
\label{tab:analyzer-precision-recall}
\setlength{\tabcolsep}{5pt}
\begin{tabular}{lrrrrrr}
\toprule
Model & N & Insecure & Flagged & Precision & Recall & F1 \\
\midrule
\multicolumn{7}{c}{\textit{BaxBench}} \\
\midrule
Qwen2.5-Coder-14B & 12 & 66.7 & 83.3 & 70.0 & 87.5 & 77.8 \\
DeepSeek-V4.1-flash & 166 & 41.0 & 34.3 & 38.6 & 32.4 & 35.2 \\
Qwen2.5-72B-Instruct & 67 & 61.2 & 37.3 & 68.0 & 41.5 & 51.5 \\
\midrule
\multicolumn{7}{c}{\textit{SecEvalPlus}} \\
\midrule
Qwen2.5-Coder-14B & 102 & 87.3 & 5.9 & 83.3 & 5.6 & 10.5 \\
DeepSeek-V4.1-flash & 347 & 62.0 & 3.5 & 75.0 & 4.2 & 7.9 \\
Qwen2.5-72B-Instruct & 280 & 81.1 & 5.4 & 80.0 & 5.3 & 9.9 \\
\midrule
\multicolumn{7}{c}{\textit{RiskyPy}} \\
\midrule
Qwen2.5-Coder-14B & 53 & 56.6 & 75.5 & 60.0 & 80.0 & 68.6 \\
DeepSeek-V4.1-flash & 211 & 45.5 & 89.6 & 47.6 & 93.8 & 63.2 \\
Qwen2.5-72B-Instruct & 181 & 58.0 & 76.8 & 61.2 & 81.0 & 69.7 \\
\bottomrule
\end{tabular}
\end{table*}

%% file: sections/results/guidance_comparaison.tex
\subsection{RQ2: Comparing Guidance Strategies}
\label{sec:rq2}

\subsubsection{Effect on Static-Analysis Findings}
\label{sec:rq2-findings}


We first examine whether guidance reduces findings from the analyzer that drives it. A lower flag rate can be misleading if broken code is no longer parseable or if the model avoids the required functionality. We therefore focus on programs that remain functional and report the share that work and are flagged by neither analyzer.


On RiskyPy, guidance produces a real gain on this measure: the best strategy for each model raises the share of working, clean programs from 10.0--19.2\% to 46.6--58.4\%. We discuss in Section~\ref{sec:rq2-security} how much this translates into more secure programs. The gain comes, however, at a cost in functionality.  Every analyzer-guided strategy lowers RiskyPy's functionality by 2.7 to 71.2 percentage points, and for some strategies more of the drop in flags comes from broken programs than from working ones. FDSP (Semgrep) removes nearly all flags on Qwen2.5-Coder-14B (74.0\% to 1.4\%), yet only 46.6\% of its programs work and are clean, while 52.0\% are unflagged but broken. PromSec is the extreme case: on Qwen2.5-Coder-14B, it lowers the flag rate to 39.7\%, but only 1.4\% of its programs work.

In BaxBench, flags drop only due to broken code. For DeepSeek-V4.1-flash, Steering and FDSP (Semgrep) cut the flag rate from 34.1\% to 3.6\% and 0.0\%, yet the share of working, clean applications barely moves (43.3\% to 44.8\% and 44.0\%). Nearly all of the removed flags were on applications that stopped working, which is what complying with \texttt{avoid\_app\_run\_with\_bad\_host} produces (Section~\ref{sec:datasets}). On SecEvalPlus, analyzers rarely fire, so there is little to remove. Semgrep-guided strategies lower flag rates of 2.9--5.0\% to 0.0--2.9\%, and the share of working, clean programs changes by at most 6.4 points.
Bandit's flag rate also falls, but this is weak evidence that guidance generalizes beyond the analyzer that drives it. The two analyzers share rules for the dominant patterns in our benchmarks (\texttt{eval}, \texttt{pickle}, \texttt{subprocess} with \texttt{shell=True}, and XML parsing), and their flag rates move together on RiskyPy.

\begin{table*}[t]
\centering
\small
\caption{SecEvalPlus results. Tests / Func. tests / Sec. tests: passes the dataset's Test / Test-FP / Test-SP unit tests. Func. + Semgrep / Bandit clean: passes Test-FP and the analyzer flags nothing; Func. + both clean: passes Test-FP and neither analyzer flags anything. Semgrep / Bandit: flagged vulnerable. All values in \%; models with several runs: mean of 3 runs. Best pass rates (higher is better) per model in bold.}
\label{tab:secevalplus}
\setlength{\tabcolsep}{4pt}
\begin{tabular}{lrrrrrrrr}
\toprule
Method
& \shortstack{Tests\\$\uparrow$}
& \shortstack{Func. +\\tests $\uparrow$}
& \shortstack{Sec. +\\tests $\uparrow$}
& \shortstack{Func. +\\Semg clean $\uparrow$}
& \shortstack{Func. +\\Band clean $\uparrow$}
& \shortstack{Func. + both\\clean $\uparrow$}
& \shortstack{Semg\\$\downarrow$}
& \shortstack{Band\\$\downarrow$} \\
\midrule
\multicolumn{9}{c}{\textit{Qwen2.5-Coder-14B}} \\
\midrule
Baseline & 9.3 & \textbf{72.9} & 10.0 & 69.3 & 68.6 & 68.6 & 5.0 & 5.7 \\
Self-Refine & 10.0 & \textbf{72.9} & 10.7 & \textbf{70.0} & \textbf{69.3} & \textbf{69.3} & 3.6 & 4.3 \\
FDSP & 9.3 & 70.7 & 11.4 & 69.3 & \textbf{69.3} & \textbf{69.3} & 2.1 & 2.1 \\
FDSP (Semgrep) & 9.3 & 70.0 & 10.7 & \textbf{70.0} & \textbf{69.3} & \textbf{69.3} & 0.0 & 0.7 \\
PromSec & 5.7 & 37.1 & 10.0 & 36.4 & 35.7 & 35.7 & 2.9 & 2.9 \\
\cmidrule(l){1-9}
Comment (ours) & \textbf{13.6} & 70.0 & \textbf{15.0} & 67.9 & 67.1 & 67.1 & 2.9 & 3.6 \\
Steering (ours) & 6.4 & 70.7 & 8.6 & \textbf{70.0} & \textbf{69.3} & \textbf{69.3} & 1.4 & 2.1 \\
\midrule
\multicolumn{9}{c}{\textit{DeepSeek-V4.1-flash (mean of 3 runs)}} \\
\midrule
Baseline & 31.4 & \textbf{82.6} & 33.3 & 79.8 & 79.8 & 79.8 & 2.9 & 2.9 \\
Self-Refine & 31.2 & 81.9 & 33.3 & \textbf{81.2} & \textbf{81.2} & \textbf{81.2} & 0.7 & 0.7 \\
FDSP & 31.2 & 81.4 & 33.8 & \textbf{81.2} & \textbf{81.2} & \textbf{81.2} & 0.2 & 0.2 \\
FDSP (Semgrep) & 30.2 & 81.2 & 33.1 & \textbf{81.2} & \textbf{81.2} & \textbf{81.2} & 0.0 & 0.0 \\
PromSec & 29.3 & 81.0 & 31.9 & 79.5 & 79.5 & 79.5 & 1.4 & 1.4 \\
\cmidrule(l){1-9}
Comment (ours) & 34.3 & 81.9 & \textbf{37.6} & 80.0 & 80.0 & 80.0 & 1.9 & 1.9 \\
Steering (ours) & \textbf{34.8} & 81.7 & 36.7 & 80.7 & 80.7 & 80.7 & 1.0 & 1.0 \\
\midrule
\multicolumn{9}{c}{\textit{Qwen2.5-72B-Instruct (mean of 3 runs)}} \\
\midrule
Baseline & 12.9 & 66.7 & 15.5 & 63.1 & 63.1 & 63.1 & 4.8 & 4.8 \\
Self-Refine & 12.4 & 65.5 & 15.5 & 65.0 & 65.0 & 65.0 & 0.5 & 0.5 \\
FDSP & 13.1 & 66.4 & 16.0 & 65.7 & 65.7 & 65.7 & 1.4 & 1.4 \\
FDSP (Semgrep) & 13.8 & 66.7 & 15.7 & 65.2 & 65.2 & 65.2 & 1.4 & 1.9 \\
PromSec & 6.4 & 46.9 & 8.3 & 44.3 & 44.3 & 44.3 & 3.3 & 3.6 \\
\cmidrule(l){1-9}
Comment (ours) & 13.8 & \textbf{70.0} & 15.7 & \textbf{69.5} & \textbf{69.5} & \textbf{69.5} & 2.6 & 3.1 \\
Steering (ours) & \textbf{14.0} & 67.1 & \textbf{16.2} & 66.2 & 66.2 & 66.2 & 1.0 & 1.0 \\
\bottomrule
\end{tabular}
\end{table*}

\begin{table*}[t]
\centering
\small
\caption{RiskyPy results. Results are reported following the same convention as in Table \ref{tab:secevalplus}.}
\label{tab:riskypy-judge}
\resizebox{\textwidth}{!}{%
\begin{tabular}{lrrrrrrrr}
\toprule
Method
& \shortstack{Works + LLM\\safe $\uparrow$}
& \shortstack{Works +\\Semg clean $\uparrow$}
& \shortstack{Works +\\Band clean $\uparrow$}
& \shortstack{Works + both\\clean $\uparrow$}
& \shortstack{Works\\$\uparrow$}
& \shortstack{LLM safe\\$\uparrow$}
& \shortstack{Semg\\$\downarrow$}
& \shortstack{Band\\$\downarrow$} \\
\midrule
\multicolumn{9}{c}{\textit{Qwen2.5-Coder-14B}} \\
\midrule
Baseline & \textbf{31.5} & 17.8 & 17.8 & 17.8 & \textbf{72.6} & 47.9 & 74.0 & 74.0 \\
Self-Refine & \textbf{31.5} & 17.8 & 17.8 & 17.8 & 71.2 & 49.3 & 74.0 & 74.0 \\
FDSP & 27.4 & 45.2 & 45.2 & 45.2 & 52.1 & 67.1 & 8.2 & 9.6 \\
FDSP (Semgrep) & 27.4 & \textbf{46.6} & \textbf{46.6} & \textbf{46.6} & 46.6 & \textbf{69.9} & 1.4 & 1.4 \\
PromSec & 1.4 & 1.4 & 1.4 & 1.4 & 1.4 & 68.5 & 39.7 & 38.4 \\
\cmidrule(lr){1-9}
Comment (ours) & \textbf{31.5} & 42.5 & 42.5 & 42.5 & 69.9 & 47.9 & 43.8 & 43.8 \\
Steering (ours) & 28.8 & 39.7 & 41.1 & 39.7 & 57.5 & 57.5 & 30.1 & 27.4 \\
\midrule
\multicolumn{9}{c}{\textit{DeepSeek-V4.1-flash (mean of 3 runs)}} \\
\midrule
Baseline & 52.5 & 10.0 & 10.0 & 10.0 & \textbf{96.3} & 54.3 & 90.0 & 90.0 \\
Self-Refine & 63.5 & 26.5 & 33.8 & 26.5 & 89.5 & 72.6 & 66.7 & 57.5 \\
FDSP & \textbf{66.2} & 53.9 & 56.6 & 53.9 & 77.6 & 86.8 & 28.8 & 24.7 \\
FDSP (Semgrep) & 61.6 & \textbf{58.4} & \textbf{58.4} & \textbf{58.4} & 71.7 & \textbf{88.6} & 19.2 & 19.2 \\
PromSec & 44.7 & 8.2 & 8.2 & 8.2 & 77.6 & 59.4 & 90.0 & 90.0 \\
\cmidrule(lr){1-9}
Comment (ours) & 45.7 & 22.4 & 22.4 & 22.4 & 81.3 & 60.3 & 72.6 & 72.6 \\
Steering (ours) & 64.4 & 57.1 & 57.1 & 57.1 & 82.2 & 77.6 & 31.1 & 31.1 \\
\midrule
\multicolumn{9}{c}{\textit{Qwen2.5-72B-Instruct (mean of 3 runs)}} \\
\midrule
Baseline & 34.7 & 19.2 & 19.2 & 19.2 & \textbf{82.6} & 42.0 & 74.9 & 74.9 \\
Self-Refine & 25.6 & 18.7 & 18.7 & 18.7 & 58.0 & 55.3 & 47.0 & 46.6 \\
FDSP & 30.6 & 47.0 & 47.0 & 47.0 & 52.1 & \textbf{69.9} & 8.2 & 8.2 \\
FDSP (Semgrep) & 30.1 & \textbf{48.9} & \textbf{48.9} & \textbf{48.9} & 55.3 & 63.9 & 9.6 & 9.6 \\
PromSec & 11.4 & 11.4 & 11.4 & 11.4 & 35.2 & 38.8 & 73.5 & 73.5 \\
\cmidrule(lr){1-9}
Comment (ours) & \textbf{35.2} & 47.9 & 47.9 & 47.9 & 71.7 & 47.9 & 44.3 & 44.3 \\
Steering (ours) & 30.6 & 44.3 & 45.2 & 44.3 & 62.6 & 56.6 & 27.4 & 25.6 \\
\bottomrule
\end{tabular}%
}
\end{table*}

\begin{table*}[t]
\centering
\small
\setlength{\tabcolsep}{4pt}
\caption{BaxBench results. Results are reported following the same convention as in Table \ref{tab:secevalplus}.}
\label{tab:baxbench}
\begin{tabular}{lrrrrrrrr}
\toprule
Method
& \shortstack{Works\\$\uparrow$}
& \shortstack{Attacked\\$\downarrow$}
& \shortstack{Works +\\safe $\uparrow$}
& \shortstack{Works +\\Semg clean $\uparrow$}
& \shortstack{Works +\\Band clean $\uparrow$}
& \shortstack{Works + both\\clean $\uparrow$}
& \shortstack{Semg\\$\downarrow$}
& \shortstack{Band\\$\downarrow$} \\
\midrule
\multicolumn{9}{c}{\textit{Qwen2.5-Coder-14B}} \\
\midrule
Baseline & 14.3 & 26.2 & 4.8 & 2.4 & 10.7 & 2.4 & 41.7 & 11.9 \\
Self-Refine & \textbf{15.5} & 25.0 & \textbf{6.0} & 3.6 & \textbf{11.9} & 3.6 & 39.3 & 8.5 \\
FDSP & 0.0 & 0.0 & 0.0 & 0.0 & 0.0 & 0.0 & 4.8 & 1.2 \\
FDSP (Semgrep) & 4.8 & 8.3 & 3.6 & 4.8 & 4.8 & 4.8 & 3.6 & 0.0 \\
PromSec & 3.6 & 4.8 & 2.4 & 0.0 & 2.4 & 0.0 & 26.2 & 3.7 \\
\cmidrule(l){1-9}
Comment (ours) & 8.3 & 13.1 & 4.8 & 6.0 & 7.1 & 6.0 & 23.8 & 15.5 \\
Steering (ours) & 8.3 & 11.9 & 3.6 & \textbf{7.1} & 8.3 & \textbf{7.1} & 16.7 & 14.3 \\
\midrule
\multicolumn{9}{c}{\textit{DeepSeek-V4.1-flash (mean of 3 runs)}} \\
\midrule
Baseline & 65.9 & 32.1 & \textbf{38.9} & 43.3 & 65.1 & 43.3 & 34.1 & 0.8 \\
Self-Refine & 65.1 & 31.3 & 37.7 & 43.7 & 65.1 & 43.7 & 33.3 & 0.0 \\
FDSP & 2.4 & 0.0 & 2.4 & 2.0 & 2.4 & 2.0 & 2.8 & 0.0 \\
FDSP (Semgrep) & 44.0 & 22.2 & 24.6 & 44.0 & 44.0 & 44.0 & 0.0 & 0.0 \\
PromSec & 61.9 & 29.5 & 38.1 & 40.5 & 61.9 & 40.5 & 33.3 & 0.4 \\
\cmidrule(l){1-9}
Comment (ours) & \textbf{68.3} & 35.3 & \textbf{38.9} & 44.4 & \textbf{67.1} & 44.4 & 35.3 & 2.0 \\
Steering (ours) & 48.0 & 22.2 & 28.6 & \textbf{44.8} & 46.8 & \textbf{44.8} & 3.6 & 1.2 \\
\midrule
\multicolumn{9}{c}{\textit{Qwen2.5-72B-Instruct (mean of 3 runs)}} \\
\midrule
Baseline & 26.6 & 33.3 & 10.3 & 16.7 & 24.2 & 16.7 & 38.1 & 7.5 \\
Self-Refine & 24.6 & 29.0 & 9.9 & 15.1 & 23.0 & 15.1 & 35.7 & 6.0 \\
FDSP & 0.0 & 0.0 & 0.0 & 0.0 & 0.0 & 0.0 & 2.4 & 0.0 \\
FDSP (Semgrep) & 20.6 & 20.6 & 8.7 & 20.6 & 20.6 & 20.6 & 0.0 & 0.0 \\
PromSec & 21.8 & 24.8 & 8.7 & 14.7 & 20.6 & 14.7 & 31.7 & 6.4 \\
\cmidrule(l){1-9}
Comment (ours) & \textbf{29.8} & 34.5 & \textbf{13.5} & \textbf{22.6} & \textbf{27.8} & \textbf{22.6} & 25.0 & 8.3 \\
Steering (ours) & 21.4 & 18.7 & 10.3 & 20.2 & 21.4 & 20.2 & 6.0 & 2.8\\
\bottomrule
\end{tabular}
\end{table*}

\subsubsection{Security and Functionality}
\label{sec:rq2-security}

Tables~\ref{tab:secevalplus}, \ref{tab:riskypy-judge}, and~\ref{tab:baxbench} report the results per benchmark.

\paragraph{Only DeepSeek-V4.1-flash benefits} On SecEvalPlus, Steering and Comment reach the highest combined rates (34.8\% and 34.3\% vs.\ 31.4\%) without affecting functionality. On RiskyPy, FDSP (66.2\%), Steering (64.4\%), and Self-Refine (63.5\%) significantly improve on the baseline (52.5\%) by trading functionality for safety: Steering raises the share judged safe from 54.3\% to 77.6\% while functionality falls from 96.3\% to 82.2\%. FDSP reaches the highest combined rate but at a much higher token cost (Section~\ref{sec:cost}), and Self-Refine nearly matches Steering without any analyzer, so much of this gain does not require static analysis. On BaxBench, no strategy beats the baseline (38.9\%). Steering lowers the share of exploited applications from 32.1\% to 22.2\%, but functionality falls from 65.9\% to 48.0\% as applications no longer start (Section \ref{subsec:overcompliance}), and its combined rate drops significantly to 28.6\%.

\paragraph{Neither Qwen model benefits} No strategy significantly improves either model's combined rate. For Qwen2.5-Coder-14B, Comment does best on SecEvalPlus (13.6\% vs.\ 9.3\%), while Steering lowers it to 6.4\%. On RiskyPy, Steering raises the share judged safe from 47.9\% to 57.5\%, but functionality falls further, from 72.6\% to 57.5\%. Its BaxBench results are too sparse to interpret, with only 14.3\% of baseline applications working. Qwen2.5-72B follows the same pattern. On SecEvalPlus, all strategies except PromSec stay within 1.1 points of the baseline. On RiskyPy, FDSP and Steering raise safety (69.9\% and 56.6\% vs.\ 42.0\%) at a larger cost in functionality (52.1\% and 62.6\% vs.\ 82.6\%), and only Comment edges past the baseline (35.2\% vs.\ 34.7\%). The larger model thus behaves like the smaller one rather than like DeepSeek-V4.1-flash (Section~\ref{sec:rq3}).

\paragraph{Iterative and prompt-optimization strategies lose the most functionality} PromSec reduces SecEvalPlus functionality from 72.9\% to 37.1\% for Qwen2.5-Coder-14B and from 66.7\% to 46.9\% for Qwen2.5-72B, and only 1.4\% and 35.2\% of their RiskyPy programs work, significant losses on both models. On DeepSeek-V4.1-flash, it stays close to the baseline but leaves RiskyPy's flag rate unchanged at 90.0\%. FDSP collapses on BaxBench (0.0\% functional for both Qwen models, 2.4\% for DeepSeek-V4.1-flash) because it rewrites the listen address. Self-Refine stays close to the baseline on SecEvalPlus and BaxBench, but on RiskyPy its effect depends on the model: negligible on Qwen2.5-Coder-14B, +11.0 points on DeepSeek-V4.1-flash, and $-$9.1 on Qwen2.5-72B, whose functionality falls from 82.6\% to 58.0\%.

\paragraph{Statistical significance.}

Table~\ref{tab:significance} reports the statistical comparison of each approach with the baseline. On the main functionality-security measure, few differences are significant. On RiskyPy, Steering (+11.9 points), FDSP (+13.7), and Self-Refine (+11.0) significantly improve DeepSeek-V4.1-flash, while PromSec worsens both Qwen models ($-30.1$ and $-23.3$). On BaxBench, all significant changes are losses, while on SecEvalPlus no approach significantly improves over the baseline.

In contrast, the improvements are much larger when security is measured through the absence of static analyzer findings. On RiskyPy, most analyzer-guided approaches significantly increase the share of working programs flagged by neither analyzer by +12.3 to +48.4 points. The difference between these two measures suggests that optimizing against static analyzers can sometimes make vulnerabilities less visible without actually fixing them. This is particularly problematic because such vulnerabilities may become harder to detect and therefore less likely to be fixed.

\begin{table*}[t]
\centering
\small
\caption{Difference from the Baseline in percentage points, on the share of tasks
that (a) work and are secure per the benchmark's oracle (SecEvalPlus: Tests;
BaxBench: Works + safe; RiskyPy: Works + LLM safe), and (b) work and are flagged
by neither analyzer. Exact McNemar test on tasks paired with the Baseline's output
for the same task and run, stratified by run for multi-run models
(Cochran--Mantel--Haenszel for matched pairs). Holm correction within each model,
benchmark, and measure. $^{*}$/$^{**}$: adjusted $p<0.05$/$0.01$. }
\label{tab:significance}
\begin{tabular}{lrrrrrrrrr}
\toprule
 & \multicolumn{3}{c}{SecEvalPlus} & \multicolumn{3}{c}{BaxBench} & \multicolumn{3}{c}{RiskyPy} \\
\cmidrule(lr){2-4}\cmidrule(lr){5-7}\cmidrule(lr){8-10}
Method & 14B & DS & 72B & 14B & DS & 72B & 14B & DS & 72B \\
\midrule
\multicolumn{10}{c}{\textit{(a) Works and secure (benchmark oracle)}} \\
Self-Refine    & $+0.7$ & $-0.2$ & $-0.5$ & $+1.2$ & $-1.2$ & $-0.4$ & $0.0$ & $+11.0^{**}$ & $-9.1$ \\
FDSP           & $0.0$ & $-0.2$ & $+0.2$ & $-4.8$ & $-36.5^{**}$ & $-10.3^{**}$ & $-4.1$ & $+13.7^{**}$ & $-4.1$ \\
FDSP (Semgrep) & $0.0$ & $-1.2$ & $+1.0$ & $-1.2$ & $-14.3^{**}$ & $-1.6$ & $-4.1$ & $+9.1$ & $-4.6$ \\
PromSec        & $-3.6$ & $-2.1$ & $-6.4^{**}$ & $-2.4$ & $-0.8$ & $-1.6$ & $-30.1^{**}$ & $-7.8$ & $-23.3^{**}$ \\
Comment (ours) & $+4.3$ & $+2.9$ & $+1.0$ & $0.0$ & $0.0$ & $+3.2$ & $0.0$ & $-6.8$ & $+0.5$ \\
Steering (ours)& $-2.9$ & $+3.3$ & $+1.2$ & $-1.2$ & $-10.3^{**}$ & $0.0$ & $-2.7$ & $+11.9^{*}$ & $-4.1$ \\
\midrule
\multicolumn{10}{c}{\textit{(b) Works and flagged by neither analyzer}} \\
Self-Refine    & $+0.7$ & $+1.4$ & $+1.9$ & $+1.2$ & $+0.4$ & $-1.6$ & $0.0$ & $+16.4^{**}$ & $-0.5$ \\
FDSP           & $+0.7$ & $+1.4$ & $+2.6$ & $-2.4$ & $-41.3^{**}$ & $-16.7^{**}$ & $+27.4^{**}$ & $+43.8^{**}$ & $+27.9^{**}$ \\
FDSP (Semgrep) & $+0.7$ & $+1.4$ & $+2.1$ & $+2.4$ & $+0.8$ & $+4.0$ & $+28.8^{**}$ & $+48.4^{**}$ & $+29.7^{**}$ \\
PromSec        & $-32.9^{**}$ & $-0.2$ & $-18.8^{**}$ & $-2.4$ & $-2.8$ & $-2.0$ & $-16.4^{**}$ & $-1.8$ & $-7.8^{**}$ \\
Comment (ours) & $-1.4$ & $+0.2$ & $+6.4^{**}$ & $+3.6$ & $+1.2$ & $+6.0^{*}$ & $+24.7^{**}$ & $+12.3^{**}$ & $+28.8^{**}$ \\
Steering (ours)& $+0.7$ & $+1.0$ & $+3.1$ & $+4.8$ & $+1.6$ & $+3.6$ & $+21.9^{**}$ & $+47.0^{**}$ & $+25.1^{**}$ \\
\bottomrule
\end{tabular}
\end{table*}

\subsubsection{Cost}
\label{sec:cost}

We measure cost as output tokens per sample relative to the unguided baseline (Table~\ref{tab:cost}a). Real-time guidance stays close to the baseline: Comment and Steering cost 0.9--1.9$\times$ across all models and benchmarks, since a rollback regenerates only part of the program. Iterative strategies cost more, and their cost grows with the number of findings. FDSP costs 1.9--2.3$\times$ on SecEvalPlus, where the analyzer rarely fires, but 4.9--9.8$\times$ on BaxBench and RiskyPy. PromSec is the most expensive (4.7--24.1$\times$). Self-Refine depends on the model, from 1.0--1.2$\times$ on Qwen2.5-Coder-14B to 14.1--18.3$\times$ on DeepSeek-V4.1-flash.

Generation time (Table~\ref{tab:cost}b) tells a partly different story. On Qwen2.5-72B, time follows tokens: real-time guidance takes 1.1--1.7$\times$ the baseline's time, against up to 20.2$\times$ for PromSec. On DeepSeek-V4.1-flash, whose baseline takes only 1.2--3.0\,s per sample, the fixed overhead of running the analyzer after each line and restarting the request dominates. Steering takes 3.4--9.6$\times$ the baseline's time for 0.9--1.7$\times$ its tokens, and is slower than FDSP~(Semgrep) on all three benchmarks. Real-time guidance is therefore the cheapest strategy in tokens, but with a fast API model, its per-line analysis can outweigh the savings in wall-clock time.

Times for Qwen2.5-Coder-14B are only indicative. The model ran a shared server and that raised biases regarding some runs. Strategies that resend a long shared prefix also benefit from prefix caching, which shortens their calls: this explains why Self-Refine takes 0.3--0.6$\times$ the baseline's time while generating 1.0--1.2$\times$ its tokens. For this model, token counts are the reliable measure of cost.




\input{tables/cost_time}

\finding{2}{No strategy is consistently superior to the Baseline. Guidance helps only the most capable model, and only when analyzer findings align with the task; when they conflict, it breaks functionality. Among analyzer-guided strategies, real-time feedback is the cheapest.}

%% file: tables/cost_time.tex

\begin{table}[t]
    \centering
    \small
    \setlength{\tabcolsep}{5pt}
    \caption{Cost of each strategy relative to the unguided baseline ($\times$\,Base).
    The Baseline row gives the absolute cost: tokens in (a), seconds in (b).
    14B: Qwen2.5-Coder-14B ; DS: DeepSeek-flash;
    72B: Qwen2.5-72B-Instruct}
    \label{tab:cost}
    \begin{tabular}{l rrr rrr rrr}
        \toprule
        & \multicolumn{3}{c}{SecEvalPlus} & \multicolumn{3}{c}{BaxBench} & \multicolumn{3}{c}{RiskyPy} \\
        \cmidrule(lr){2-4} \cmidrule(lr){5-7} \cmidrule(lr){8-10}
        Method & 14B & DS & 72B & 14B & DS & 72B & 14B & DS & 72B \\
        \midrule
        \multicolumn{10}{c}{\textit{(a) Tokens ($\times$ Base; Baseline row: tokens per sample)}} \\
        Baseline & 144 & 167 & 125 & 546 & 799 & 566 & 363 & 445 & 364 \\
        Self-Refine & 1.1 & 15.0 & 5.9 & 1.0 & 14.1 & 10.9 & 1.2 & 18.3 & 9.1 \\
        FDSP & 1.9 & 2.3 & 2.3 & 4.9 & 6.5 & 5.6 & 7.0 & 9.8 & 7.8 \\
        FDSP (Semgrep) & 1.3 & 1.3 & 1.5 & 2.4 & 2.7 & 2.4 & 4.5 & 8.5 & 5.4 \\
        PromSec & 6.7 & 7.9 & 10.3 & 4.7 & 19.6 & 19.0 & 9.6 & 24.1 & 21.9 \\
        Comment (ours) & 1.4 & 1.4 & 1.5 & 1.1 & 0.9 & 1.1 & 1.4 & 1.9 & 1.3 \\
        Steering (ours) & 1.5 & 1.2 & 1.4 & 1.1 & 0.9 & 1.0 & 1.4 & 1.7 & 1.2 \\
        \midrule
        \multicolumn{10}{c}{\textit{(b) Generation time ($\times$ Base; Baseline row: seconds per sample)}} \\
        Baseline & 14.2 & 1.2 & 8.6 & 125.7 & 3.0 & 30.2 & 69.6 & 2.0 & 20.2 \\
        Self-Refine & 0.6 & 10.5 & 4.5 & 0.3 & 14.3 & 9.2 & 0.4 & 18.0 & 8.4 \\
        FDSP & 1.1 & 1.8 & 3.2 & 1.3 & 5.7 & 6.0 & 2.1 & 8.0 & 7.9 \\
        FDSP (Semgrep) & 0.8 & 1.4 & 1.2 & 0.6 & 2.5 & 1.9 & 1.4 & 7.0 & 4.7 \\
        PromSec & 4.0 & 6.4 & 7.3 & 1.2 & 19.9 & 16.9 & 3.1 & 23.3 & 20.2 \\
        Comment (ours) & 1.9 & 2.4 & 1.3 & 1.3 & 3.6 & 1.1 & 1.9 & 5.4 & 1.4 \\
        Steering (ours) & 2.1 & 3.4 & 1.6 & 1.2 & 4.8 & 1.1 & 1.9 & 9.6 & 1.7 \\
        \bottomrule
    \end{tabular}
\end{table}

%% file: sections/results/drive_success.tex
\subsection{RQ3: What Drives Success and Failures}
\label{sec:rq3}

\subsubsection{Model}
\label{sec:rq3-model}

The model determines the sign of Steering's effect more than any other factor. Table~\ref{tab:net-rule} reports, for each rule, the share of sample-rule pairs that Steering solved minus the share it broke. On DeepSeek-flash, the net effect is positive on three of the five rules that interrupt generation on RiskyPy: \texttt{eval-detected} ($+63.6$, $n=33$), \texttt{use-defused-xml} ($+28.3$, $n=99$), and \texttt{avoid-pickle} ($+41.7$, $n=12$), and close to zero on \texttt{dynamic-urllib-use-detected} ($-4.1$, $n=73$) and \texttt{subprocess-shell-true} ($-2.2$, $n=45$). On both Qwen models, the same rules are net-negative, with \texttt{eval-detected} at $-31.6$ (Qwen2.5-Coder-14B) and $-33.3$ (Qwen2.5-72B), and \texttt{avoid-pickle} at $-50.0$ and $-21.4$. The only rule on which a Qwen model gains is \texttt{use-defused-xml}, and only Qwen2.5-72B ($+3.0$, against $-6.1$ for Qwen2.5-Coder-14B).

Model size does not substantially change this picture. On RiskyPy, the benchmark where most pairs occur, Qwen2.5-Coder-14B and Qwen2.5-72B reach similar outcomes (Solved 1.5\% and 6.5\%, Broke 21.2\% and 16.9\%), while DeepSeek-flash solves 25.0\% of its pairs and breaks 5.5\%. The Qwen models also leave the finding in place more often (the rule still fires in 31.8\% and 34.8\% of pairs, against 25.8\%). A fivefold increase in parameters within the Qwen2.5 family thus brings Steering no closer to its effect on DeepSeek-flash.

The rules on which the Qwen models succeed or fail suggest why. \texttt{use-defused-xml} can be addressed by a local, mechanical change: replacing the standard XML parser with its \texttt{defusedxml} counterpart. \texttt{eval-detected} and \texttt{subprocess-shell-true} usually cannot: removing \texttt{eval} from an expression evaluator, or turning a shell string into an argument list, requires restructuring the surrounding code. When the Qwen models remove the flagged pattern, they often fail to rebuild a working program around its replacement; the replication package includes examples of such Broke pairs for \texttt{eval-detected}. DeepSeek-flash handles both kinds of fix, except on \texttt{subprocess-shell-true}, which is net-negative on all three models.

\finding{3}{The model sets the sign of the effect. Steering is net-positive on half of the rules with ten or more pairs for DeepSeek-flash, but net-negative on all of them except use-defused-xml for Qwen2.5-72B, and on all of them for Qwen2.5-Coder-14B. The Qwen models apply mechanical fixes but often break the program when a fix requires restructuring it, regardless of size.}

\subsubsection{Rule Type and Severity}
\label{sec:rq3-rule}
Only 11 of the 225 rules interrupted Steering, and only six did so on ten or more pairs for any model. Two of these six are ERROR rules, and they have opposite outcomes: use-defused-xml is the only rule net-positive on more than one model, while subprocess-shell-true is net-negative on all three (Table~\ref{tab:net-rule}). Severity does not seem to predict the outcome.

Within this set, outcomes follow what a fix requires. \rulename{use-defused-xml} needs only a parser substitution and is net-positive on two of the three models. \rulename{eval-detected}, \rulename{avoid-pickle}, and \rulename{subprocess-shell-true} require restructuring the surrounding code and help only DeepSeek-flash, \rulename{subprocess-shell-true} excepted (Section~\ref{sec:rq3-model}). Finally, some rules flag behavior the task requires, so no fix removes the finding without changing the program. \rulename{avoid_app_run_with_bad_host} is never solved and breaks up to 25.9\% of pairs, while \rulename{dynamic-urllib-use-detected}, which flags requests to user-supplied URLs, survives Steering in 53.4\% of DeepSeek-flash's pairs. 
For such rules, leaving the finding in place is the desirable outcome.

\finding{4}{Few rules drive Steering, and severity does not seem to predict effectiveness. Substitutions succeed on two of the three models, restructuring succeeds mainly on the stronger model, and findings that contradict the task either break the program or are rightly ignored.}

\subsubsection{Overcompliance}
\label{subsec:overcompliance}

BaxBench requires the server to listen on 0.0.0.0 so that tests and exploits can reach it from outside the container, yet Semgrep's \texttt{avoid\_app\_run\_with\_bad\_host} flags exactly this required binding. The finding is therefore a false positive in context, and although Steering asks the LLM to follow a recommendation only when it does not conflict with the task, the required binding was preserved in only 4 of the 83 pairs where the rule fired and the baseline was functional. Two of the three models never preserved it, the rule was never resolved, and Steering broke 15.9-25.9\% of pairs depending on the LLM, with the remaining pairs labeled Did nothing or Already broken.

On other rules, however, compliance proved to be less unconditional. On dynamic-urllib-use-detected, which flags requests to runtime-computed URLs that many RiskyPy tasks describe, DeepSeek-V4.1-flash keeps the flagged behavior in 53.4\% of pairs. Models can decline a finding, but not reliably, even when told to prioritize the specification.

\finding{5}{Models overcomply with findings that contradict the specification. When told to ignore conflicting recommendations, they still removed a deployment-required binding in 79 of 83 cases.}

\begin{table}[t]
\centering
\small
\caption{Net effect of Steering per rule: Solved minus Broke, in percentage points of the rule's $n$ ($n$ in parentheses), pooled over SecEvalPlus, BaxBench, and RiskyPy (per-benchmark counts in the replication package). DS and 72B pool three runs; pairs from the same task are not independent. --: rule never interrupted Steering. $^\ddagger n<10$.}
\label{tab:net-rule}
\begin{tabular}{l l r r r}
\toprule
Rule & Severity & Qwen2.5-Coder-14B & DeepSeek-flash & Qwen2.5-72B \\
\midrule
eval-detected & WARNING & $-31.6$ (19) & $+63.6$ (33) & $-33.3$ (63) \\
use-defused-xml & ERROR & $-6.1$ (33) & $+28.3$ (99) & $+3.0$ (99) \\
avoid-pickle & WARNING & $-50.0$ (6)$^\ddagger$ & $+41.7$ (12) & $-21.4$ (14) \\
dynamic-urllib-use-detected & WARNING & -- & $-4.1$ (73) & -- \\
exec-detected & WARNING & $0.0$ (1)$^\ddagger$ & $0.0$ (3)$^\ddagger$ & $0.0$ (3)$^\ddagger$ \\
render-template-string & WARNING & $0.0$ (1)$^\ddagger$ & $0.0$ (8)$^\ddagger$ & $0.0$ (4)$^\ddagger$ \\
secure-set-cookie & WARNING & $0.0$ (1)$^\ddagger$ & $0.0$ (3)$^\ddagger$ & $0.0$ (3)$^\ddagger$ \\
subprocess-shell-true & ERROR & $-27.8$ (18) & $-2.2$ (45) & $-4.2$ (48) \\
avoid\_app\_run\_with\_bad\_host & WARNING & $-25.9$ (27) & $-23.1$ (78) & $-15.9$ (82) \\
dangerous-template-string & ERROR & -- & -- & $0.0$ (3)$^\ddagger$ \\
eval-injection & ERROR & -- & -- & $0.0$ (1)$^\ddagger$ \\
\bottomrule
\end{tabular}
\end{table}

\begin{table*}[t]
\centering
\caption{Outcomes of Steering per rules: share of sample-rule pairs (count in parentheses). Net: Solved minus Broke, in percentage points. DeepSeek-flash and Qwen2.5-72B pool three runs; pairs from the same task are not independent. $^\ddagger n<10$. Per-benchmark counts are in the replication package.}
\label{tab:rule-outcomes}
\scriptsize
\begin{tabular}{l r r r r r r r}
\toprule
\textbf{Rule} & \textbf{$n$} & \textbf{Solved} & \textbf{Broke} & \twoline{Did}{nothing} & \twoline{Already}{broken} & \textbf{Silenced} & \textbf{Net} \\
\midrule
\multicolumn{8}{c}{\textbf{Qwen2.5-Coder-14B}} \\
\midrule
\texttt{use-defused-xml} & 33 & 0.0\% (0) & 6.1\% (2) & 69.7\% (23) & 3.0\% (1) & 21.2\% (7) & $-6.1$ \\
\texttt{avoid\_app\_run\_with\_bad\_host} & 27 & 0.0\% (0) & 25.9\% (7) & 51.9\% (14) & 22.2\% (6) & 0.0\% (0) & $-25.9$ \\
\texttt{eval-detected} & 19 & 0.0\% (0) & 31.6\% (6) & 57.9\% (11) & 5.3\% (1) & 5.3\% (1) & $-31.6$ \\
\texttt{subprocess-shell-true} & 18 & 0.0\% (0) & 27.8\% (5) & 50.0\% (9) & 16.7\% (3) & 5.6\% (1) & $-27.8$ \\
\texttt{avoid-pickle}$^\ddagger$ & 6 & 16.7\% (1) & 66.7\% (4) & 0.0\% (0) & 16.7\% (1) & 0.0\% (0) & $-50.0$ \\
\texttt{exec-detected}$^\ddagger$ & 1 & 0.0\% (0) & 0.0\% (0) & 100.0\% (1) & 0.0\% (0) & 0.0\% (0) & $0.0$ \\
\texttt{render-template-string}$^\ddagger$ & 1 & 0.0\% (0) & 0.0\% (0) & 100.0\% (1) & 0.0\% (0) & 0.0\% (0) & $0.0$ \\
\texttt{secure-set-cookie}$^\ddagger$ & 1 & 0.0\% (0) & 0.0\% (0) & 100.0\% (1) & 0.0\% (0) & 0.0\% (0) & $0.0$ \\
\midrule
\textbf{ALL} & 106 & 0.9\% (1) & 22.6\% (24) & 56.6\% (60) & 11.3\% (12) & 8.5\% (9) & $-21.7$ \\
\midrule
\multicolumn{8}{c}{\textbf{DeepSeek-flash (3 runs pooled)}} \\
\midrule
\texttt{use-defused-xml} & 99 & 30.3\% (30) & 2.0\% (2) & 54.5\% (54) & 3.0\% (3) & 10.1\% (10) & $+28.3$ \\
\texttt{avoid\_app\_run\_with\_bad\_host} & 78 & 0.0\% (0) & 23.1\% (18) & 71.8\% (56) & 5.1\% (4) & 0.0\% (0) & $-23.1$ \\
\texttt{dynamic-urllib-use-detected} & 73 & 2.7\% (2) & 6.8\% (5) & 80.8\% (59) & 4.1\% (3) & 5.5\% (4) & $-4.1$ \\
\texttt{subprocess-shell-true} & 45 & 8.9\% (4) & 11.1\% (5) & 73.3\% (33) & 0.0\% (0) & 6.7\% (3) & $-2.2$ \\
\texttt{eval-detected} & 33 & 63.6\% (21) & 0.0\% (0) & 18.2\% (6) & 0.0\% (0) & 18.2\% (6) & $+63.6$ \\
\texttt{avoid-pickle} & 12 & 58.3\% (7) & 16.7\% (2) & 0.0\% (0) & 0.0\% (0) & 25.0\% (3) & $+41.7$ \\
\texttt{render-template-string}$^\ddagger$ & 8 & 0.0\% (0) & 0.0\% (0) & 100.0\% (8) & 0.0\% (0) & 0.0\% (0) & $0.0$ \\
\texttt{exec-detected}$^\ddagger$ & 3 & 0.0\% (0) & 0.0\% (0) & 33.3\% (1) & 0.0\% (0) & 66.7\% (2) & $0.0$ \\
\texttt{secure-set-cookie}$^\ddagger$ & 3 & 0.0\% (0) & 0.0\% (0) & 100.0\% (3) & 0.0\% (0) & 0.0\% (0) & $0.0$ \\
\midrule
\textbf{ALL} & 354 & 18.1\% (64) & 9.0\% (32) & 62.1\% (220) & 2.8\% (10) & 7.9\% (28) & $+9.0$ \\
\midrule
\multicolumn{8}{c}{\textbf{Qwen2.5-72B-Instruct (3 runs pooled)}} \\
\midrule
\texttt{use-defused-xml} & 99 & 9.1\% (9) & 6.1\% (6) & 65.7\% (65) & 2.0\% (2) & 17.2\% (17) & $+3.0$ \\
\texttt{avoid\_app\_run\_with\_bad\_host} & 82 & 0.0\% (0) & 15.9\% (13) & 69.5\% (57) & 14.6\% (12) & 0.0\% (0) & $-15.9$ \\
\texttt{eval-detected} & 63 & 3.2\% (2) & 36.5\% (23) & 50.8\% (32) & 7.9\% (5) & 1.6\% (1) & $-33.3$ \\
\texttt{subprocess-shell-true} & 48 & 2.1\% (1) & 6.2\% (3) & 79.2\% (38) & 12.5\% (6) & 0.0\% (0) & $-4.2$ \\
\texttt{avoid-pickle} & 14 & 21.4\% (3) & 42.9\% (6) & 0.0\% (0) & 21.4\% (3) & 14.3\% (2) & $-21.4$ \\
\texttt{render-template-string}$^\ddagger$ & 4 & 0.0\% (0) & 0.0\% (0) & 75.0\% (3) & 25.0\% (1) & 0.0\% (0) & $0.0$ \\
\texttt{exec-detected}$^\ddagger$ & 3 & 0.0\% (0) & 0.0\% (0) & 100.0\% (3) & 0.0\% (0) & 0.0\% (0) & $0.0$ \\
\texttt{secure-set-cookie}$^\ddagger$ & 3 & 0.0\% (0) & 0.0\% (0) & 100.0\% (3) & 0.0\% (0) & 0.0\% (0) & $0.0$ \\
\texttt{dangerous-template-string}$^\ddagger$ & 3 & 0.0\% (0) & 0.0\% (0) & 66.7\% (2) & 33.3\% (1) & 0.0\% (0) & $0.0$ \\
\texttt{eval-injection}$^\ddagger$ & 1 & 0.0\% (0) & 0.0\% (0) & 100.0\% (1) & 0.0\% (0) & 0.0\% (0) & $0.0$ \\
\midrule
\textbf{ALL} & 320 & 4.7\% (15) & 15.9\% (51) & 63.8\% (204) & 9.4\% (30) & 6.2\% (20) & $-11.2$ \\
\bottomrule
\end{tabular}
\end{table*}


%% file: sections/results/failure_mode.tex

%% file: sections/related_works.tex


%% file: sections/discussion.tex
\section{Discussion and Implications}
\label{sec:discussion}


Our results echo problems the software engineering community has long had with static analyzers: they overfire, have limited scope, seeing only a subset of issues, and produce warnings that are hard to act on. However, the implications of each limitation change with an LLM as recipient.

\paragraph{From fatigue to overcompliance.}
Developers respond to false positives by ignoring warnings \cite{johnson2013why, christakis2016developers}, which leaves their code intact. Models tend to follow the analyzer instead. On RiskyPy, 39--52\% of the baseline's flagged functional programs are judged secure, so about half of the flags are false alarms. When the task requires the flagged behavior, following the alert breaks the program: on BaxBench, the models removed the required network binding in 79 of 83 cases, although the prompt instructed them to give priority to the task. This damage can escape the loop, since a server bound to a loopback address fails only once deployed. \emph{Implication:} alerts on behavior that the specification or deployment requires should be filtered before they reach the model; an instruction to ignore them does not suffice. The replication package includes examples of overcompliance.

\paragraph{From blind spot to ceiling.}
Analyzers miss logic, authentication, and configuration flaws \cite{zheng2006value}. Developers compensate with review and testing, but in a guidance loop the analyzer's blind spot becomes the system's. On SecEvalPlus, the analyzers flag at most 5.6\% of the functional but insecure programs, and on RiskyPy, 8.6--12.1\% of steered pairs clear the alert but remain insecure. \emph{Implication:} guidance can only fix what the analyzer reports, so it complements review and security testing rather than replacing them.

\paragraph{Analyzer-Clean Is Often Not Secure}  Table \ref{tab:significance} scores the same programs against two definitions of success. On RiskyPy, four analyzer-guided strategies (both versions of FDSP, Comment, and Steering) significantly increase the share of working, analyzer-clean programs on all three models, by 12.3 to 48.4 points (12 of 12 comparisons significant). Against the independent security oracle, only 2 of these 12 comparisons are significant, both on DeepSeek-V4.1-flash (FDSP +13.7, Steering +11.9), and even when the guidance works, the analyzer remains an overly optimistic metric. Evaluating analyzer-guided generation with the guiding analyzer therefore overstates its effect.


\paragraph{Acting on a finding requires capability.} Acting on a warning requires understanding the code around it \cite{christakis2016developers}. Both Qwen2.5 models, at 14B and 72B, lose on rules whose fix requires restructuring, where DeepSeek-flash mostly gains, consistent with prior work on model-generated hints \cite{yan2025guiding}. \emph{Implication:} reserve analyzer guidance for capable models, and apply it during generation, at about $2\times$ the baseline's tokens rather than up to $24\times$.

\paragraph{A few rules earn their place.}
Like industrial deployments that keep only precise, actionable analyses \cite{sadowski2015tricorder}, our results favor rules with local fixes: \texttt{use-defused-xml} is the only rule net-positive on more than one model, though nine rules allow no firm conclusion. \emph{Implication:} forward only rules with a local, unambiguous fix.

\paragraph{Outlook.}
Static-analysis guidance inherits every known weakness of static analysis, and LLMs turn the most familiar one, false positives, into severe functionality damage. A promising direction is to treat findings as claims the model may contest, requiring it to either fix a finding or justify, against the specification, why the flagged behavior is needed.

%% file: sections/threath_of_validity.tex
\section{Threats to Validity}
\label{sec:threats}

\paragraph{Construct validity.}
On RiskyPy, an LLM judge assesses functionality and security. The benchmark authors validated the functionality judge, which agreed with the majority vote of three experts on 83.3\% of 30 generations; we also validate the security judge against two annotators (Section~\ref{sec:human-eval}). The security oracles in other benchmarks may also miss vulnerabilities, as tests can have gaps in coverage \cite{wang2024your}. Finally, cost counts output tokens only. Iterative strategies resend the code with each call, so counting input tokens would increase their cost relative to real-time guidance.

\paragraph{Internal validity.}
 Our reimplementations of FDSP and PromSec deviate from the original methods where needed to run them in a shared harness (Section \ref{sec:adjustments}). We document all deviations in the replication package, and where we had a choice, we resolved it in favor of the reimplemented method (e.g., returning the best-scoring rather than the last round). All strategies use the same untuned, neutral system prompt; alternative prompts may affect absolute results. Steering's recommendations add knowledge beyond the analyzer's message. Although they are written once per rule from rule metadata before the experiments, they remain a potential confound. Comment, which receives only the analyzer's message, partly controls for this. We include no prompt-only security baseline: Self-Refine, which uses no analyzer, is the closest control.
The Qwen2.5 models differ from DeepSeek-V4.1-flash in family, release date, and size, so we cannot separate capability from these factors.
All experiments use greedy decoding, and the API models are averaged over only three runs (Section \ref{sec:models}), so results under sampling may differ and run-to-run variance is only partly captured.

\paragraph{External validity.}
We use two analyzers, Semgrep and Bandit, with a frozen, untuned rule set. Prior work uses the same analyzers, and real-time guidance needs analysis fast enough to run after every line, which rules out database-based analyzers such as CodeQL; more precise data-flow analysis could change our results. Our benchmarks are Python-only, but few offer a security oracle independent of static analysis, and ours span the alignment spectrum. We ran models from only two families and only 11 rules, six with ten or more pairs, interrupted Steering. Finally, we study black-box guidance on single snippets, the setting of the approaches we compare, and small and medium models without agent scaffolding; agents may react differently to findings.

\paragraph{Conclusion validity.}
Per-rule samples are small, and pairs from the same task and pooled runs are not independent.
Differences from the baseline are tested with McNemar's test on same-task outcomes.

%% file: sections/conclusion.tex
\section{Conclusion}
\label{sec:conclusion}

In this paper, we studied whether static analysis can guide LLMs toward more secure code. We compared seven strategies across three models and three benchmarks spanning the static-analysis alignment spectrum. Overall, the results are mixed. Three problems limit what static-analysis guidance can achieve. First, its scope is narrow: on SecEvalPlus, analyzers flag about 5\% of insecure programs, and even on RiskyPy, which was built from analyzer flags, they miss 6–20\% of insecure programs. Second, it depends on the model's capability. Both Qwen2.5 models remove the flagged pattern but often break the program when the fix requires restructuring, regardless of size. Third, we noticed that models tend to overcomply with the firings they receive. When the task requires the behavior a rule flags, they tend to side with the analyzer, even when the prompt tells them to prioritize the specification, and break the program as a result.


Still, there are some positive results. First, in some instances, models correctly refused to overcomply when the finding conflicted with the task. For instance, on dynamic-urllib-use-detected, DeepSeek-V4.1-flash often kept the flagged behavior instead of breaking the program. Some rules are also net-positive. use-defused-xml improves outcomes on two of the three models, and eval-detected and avoid-pickle do so on DeepSeek-V4.1-flash. Finally, guidance does not have to be expensive. Real-time guidance costs 0.9–1.9× the baseline's tokens, far less than the iterative approaches.

These results point to two directions for future research. The first is to make analyzers fire less often and more precisely, so that fewer false alarms reach the model. The second is to help models respond to findings in context, for example by letting them either fix a finding or justify, against the specification, why the flagged behavior is needed.

For practitioners, static-analysis guidance has its benefits, but it is not a drop-in solution. Using it successfully requires choosing which rules to forward, filtering findings that conflict with the deployment, and using a sufficiently capable model. The literature does not yet offer clear guidance on how to do this.
\section*{Data Availability}
The data and code used in this study are publicly available in our
\href{https://github.com/ChL-Z/LLM-SAD.git}{ replication package}.
